\documentclass[useAMS,usenatbib]{mn2e}

\usepackage[T1]{fontenc}
\usepackage{ae,aecompl}
\usepackage{graphicx}
\usepackage{natbib}
\usepackage{amsmath}
\usepackage{amssymb}
\usepackage{multirow}
\usepackage{tabularx}
\usepackage{hyperref}
\usepackage{color}
\usepackage{ulem}

\title[Random Samples : DR12 data and QPM mocks]{Smoothing the redshift distributions of random samples for the baryon acoustic oscillations : applications to the SDSS-III BOSS DR12 and QPM mock samples}

\author[Shao-Jiang Wang, Qi Guo \& Rong-Gen Cai]{
Shao-Jiang Wang,$^{1,3,5}$\thanks{E-mail: schwang@itp.ac.cn}
Qi Guo,$^{2,4}$
Rong-Gen Cai,$^{1,3,5}$
\\
\\
$^{1}$ CAS Key Laboratory of Theoretical Physics, Institute of Theoretical Physics, Chinese Academy of Sciences, \\ \quad No.55 Zhong Guan Cun East Road, Beijing 100190, P.R. China\\
$^{2}$ National Astronomical Observatories, Chinese Academy of Sciences,\\ \quad 20A Datun Road, Chaoyang, Beijing 10012, P.R. China\\
$^{3}$ School of Physical Sciences, University of Chinese Academy of Sciences, \\ \quad No.19A Yuquan Road, Beijing 100049, P.R. China\\
$^{4}$ School of Astronomy and Space Science, University of Chinese Academy of Sciences, \\ \quad No.19A Yuquan Road, Beijing 100049, P.R. China\\
$^{5}$ Center for Gravitational Physics, Yukawa Institute for Theoretical Physics, \\ \quad Kyoto University, Kyoto 606-8502, Japan
}

\date{Accepted XXX. Received YYY; in original form ZZZ}

\pubyear{2017}

\begin{document}
\label{firstpage}
\pagerange{\pageref{firstpage}--\pageref{lastpage}}
\maketitle

\begin{abstract}
  We investigate the impact of different redshift distributions of random samples on the baryon acoustic oscillations (BAO) measurements of $D_V(z)r_\mathrm{d}^\mathrm{fid}/r_\mathrm{d}$ from the two-point correlation functions (2PCF) of galaxies in the Data Release 12 (DR12) of the Baryon Oscillation Spectroscopic Survey (BOSS). Big surveys, such as BOSS, usually assign redshifts to the random samples by randomly drawing values from the measured redshift distributions of the data, which would necessarily introduce fiducial signals of fluctuations into the random samples, weakening the signals of BAO, if the cosmic variance cannot be ignored. We propose a smooth function of redshift distribution that fits the data well to populate the random galaxy samples. The resulting cosmological parameters match the input parameters of the mock catalogue very well. The significance of BAO signals has been improved by $0.33\sigma$ for a low-redshift (LOWZ) sample and by $0.03\sigma$ for a constant-stellar-mass (CMASS) sample, though the absolute values do not change significantly. Given the precision of the measurements of current cosmological parameters, it would be appreciated for the future improvements on the measurements of galaxy clustering.
\end{abstract}

\begin{keywords}
galaxies: structure,
galaxies: statistics,
cosmology: observations,
(cosmology:) distance scale,
(cosmology:) large-scale structure of Universe
\end{keywords}

\section{Introduction}\label{sec:introduction}

The baryon acoustic oscillations (BAO) is an important tool in modern cosmology. It is the relic imprint of the matter perturbations when the sound waves stop propagating in the baryon-photon fluid after the recombination of the Universe. It is also a well-understood linear-theory phenomenon that usually serves as the standard ruler for the large-scale measurements of our Universe. When combined with other cosmological measurements, such as the Type Ia supernovae, gravitational lensing, cosmic microwave background (CMB), etc., it allows us to explore the expansion history of the whole Universe and to constrain the cosmological parameters to a high precision level.

The measurements of BAO have made great progress over the last decade. From the seminal works of the Two-Degree Field Galaxy Redshift Survey~\citep[2dFGRS;][]{Cole:2005sx} and the Sloan Digital Sky Survey~\citep[SDSS;][]{Eisenstein:2005su}, followed by enlarged measurements of the WiggleZ survey~\citep{Blake:2011en} and the Six-Degree Field Galaxy Survey~\citep[6dFGRS;][]{Beutler:2011hx}, the measurements of BAO finally reached the milestone of 1 percent precision in the SDSS-III Baryon Oscillation Spectroscopic Survey (BOSS) Data Release 11~\citep[DR11;][]{Anderson:2013zyy}, which was consistent with the results of recent Data Release 12~\citep[DR12;][]{Alam:2015mbd} of SDSS-III BOSS. In the final data release~\citep{Alam:2016hwk} of SDSS-III BOSS, the measurements of BAO were carried out using a combined sample consisting of the CMASS (constant-stellar-mass), LOWZ (low-redshift), LOWZ2 and LOWZ3 samples.

Anisotropy features have been found in BAO signals along radial and transverse directions, which arise from two main effects as follows. One is from the late-time non-linear evolutions of galaxy clustering. \cite{Eisenstein:2006nk,Padmanabhan:2012hf} proposed to reconstruct the linear power spectrum to improve the BAO signals. The other is from both the redshift-space distortions (RSD)~\citep{Kaiser:1987qv} and the mismatched fiducial cosmology, namely the Alcock-Paczynski effect~\citep{Alcock:1979mp}. The degeneracy between the \textit{Hubble} parameter $H(z)$ and the angular diameter distance $D_A(z)$ can be broken by projecting a galaxy clustering correlation along radial/transverse directions~\citep{Okumura:2007br,Gaztanaga:2008xz,Blake:2011ep,Chuang:2011fy}, or monopole/quadrupole moments~\citep{Padmanabhan:2008ag,Taruya:2011tz,Chuang:2012ad,Xu:2012fw} or the newly proposed wedge method~\citep{Kazin:2011xt,Kazin:2013rxa}. More detailed discussions on the systematics of BAO can be found in~\cite{Vargas-Magana:2016imr}.

The two point correlation function (2PCF) of galaxies is one of the most popular method to estimate the BAO signals. It measures the pairwise excesses of the data galaxies in comparison with otherwise randomly distributed galaxies. The redshift distribution $N(z)$ for the random galaxy catalogues is supposed to be smooth, given the relatively slow evolution of the tracer galaxies. However, if the volume is small, the structures along line-of-sight direction usually manifest themselves as violent fluctuations in $N(z)$. If one assumes the redshift distributions of random samples to be the same as the data samples, it would bias the resulting clustering measurements. In the literatures, many works adopted a smoothed fitting function to describe the underlying $N(z)$. Shaun Cole~\citep{Cole:2011zh} developed a new algorithm to generate the random catalogue from the data catalogue consisting of a simple flux-limited sample. It automatically provides the distribution of an unbiased $N(z)$ without assuming a prior fitting formula. For a large galaxy survey, the volume is usually big enough so that the cosmic variance could be minor.
In this case, redshifts of random galaxies are usually generated by randomly drawing from the measured galaxy redshifts with some weights
\citep[e.g. see][section 5.2]{Reid:2015gra}. Whether a random catalogue generated in this way could inherit fluctuations of the clustering signals and lead to a weakened BAO signal depends on the volume of the survey. \cite{Ross:2012qm} demonstrated the systematic difference of spherically averaged correlation functions calculated with this redshift assignment scheme and that with `spline' approaches are negligible. However, their typical spline method adopted $\Delta z=0.01$ to construct the random catalogue corresponding to $80 \sim 90\,\mathrm{Mpc}$, which is smaller than the typical scale of BAO, suggesting that the random catalogue generated using this spline approach could also inherit the BAO signals along the redshift direction.

In this paper, we propose to use a new smoothed redshift distribution function for the random galaxy samples. We describe the data and mock catalogues in Section \ref{subsec:datamock}, review the correlation function in Section \ref{subsec:correlation} and the fitting template in Section \ref{subsec:fitting}. In Section \ref{sec:methodology}, we describe how we generate the random catalogues with a wiggly redshift distribution in Section \ref{subsec:wigglymethod} and with a smoothed redshift distribution in Section \ref{subsec:smoothmethod}. The method validation is given in Section \ref{subsec:mock} using QPM mock catalogues~\citep{White:2013psd}. In Section \ref{sec:data}, we apply the smooth method to the DR12 data catalogues and compare with results using a wiggly method. Section \ref{sec:conclusions} is devoted to the conclusions.

The fiducial cosmology we used in this paper is the same as those used in the SDSS-III BOSS DR12~\citep[e.g.][]{Cuesta:2015mqa}: $h=0.70$, $\Omega_\mathrm{m}=0.29$, $\Omega_\Lambda=0.71$, $\Omega_\mathrm{b}h^2=0.02247$, $n_s=0.97$, $\sigma_8=0.80$ and $T_0=2.7255\,\mathrm{K}$. Note that we apply our method to the pre-reconstruction data and thus calculate only the monopole of the clustering. The estimate of multipoles requires the reconstructed data which will be presented in future works.

\section{Data and Method}\label{sec:formalism}

In this section, we will first describe the data and mock catalogues used in our study and then review the essential formalism to extract the signals of BAO from the observed galaxies positions.

\subsection{Data and Mocks}\label{subsec:datamock}

We use the public data catalogues from SDSS-III BOSS DR12~\citep{Alam:2015mbd}, which consist of LOWZ and CMASS samples that occupy $3.7 \mathrm{Gpc}^3$ and $10.8 \mathrm{Gpc}^3$, respectively. The LOWZ sample has totally $361,762$ galaxies covering the redshift range $0.15<z<0.43$, of which $248,237$ galaxies are from the North Galactic Cap and  $113,525$ galaxies are from the South Galactic Cap. The CMASS sample has total $777\,202$ galaxies covering the redshift range $0.43<z<0.70$, of which $568\,776$ galaxies are from the North Galactic Cap and  $208,426$ galaxies are from the South Galactic Cap. The total number of data galaxies we used for our analysis is $1,138,964$.

The random galaxy catalogues corresponding to the DR12 data galaxy catalogues are generated by the \textsc{MKSAMPLE} code, as descried in~\cite{Reid:2015gra}, to reproduce the geometry, redshift distribution and completeness of the survey. In these random samples, the redshift distribution is the same as that in the real data. In Section \ref{sec:methodology}, we will discuss two different kinds of redshift assignments of random catalogues.

Since we have only one Universe to observe, simulations of the observed galaxy clustering are crucial for determining the errors of the measurements. The mock catalogues were generated for this purpose to calculate the covariance matrix for the measured 2PCF. The QPM mock catalogues~\citep{White:2013psd} are generated using rapid, low-resolution particle mesh simulations, which mimic the observed data on both the angular selection function and the redshift distribution as well as the noise level. The fiducial cosmology is the same as we used in this paper. We use 1000 QPM mock catalogues for each sample (LOWZ or CMASS). The mock catalogues can be used not only to estimate the errors, but also to test the reliability of certain methods. We will test our method of generating random sample with the smooth redshift distribution with the QPM mock catalogues in Section \ref{subsec:mock}.

\subsection{Correlation Function}\label{subsec:correlation}

The 2PCF estimates the galaxy clustering by counting the excess of data-data galaxy pairs relative to those of random-random galaxy pairs. Here, we use the Landy and Szalay estimator~\citep{Landy:1993yu},
\begin{equation}\label{eq:xi}
\xi(r)=\frac{DD(r)-2DR(r)+RR(r)}{RR(r)},
\end{equation}
where $DD(r)$, $DR(r)$ and $RR(r)$ are the normalized number count of data-data, data-random and random-random galaxy pairs, respectively. This estimator could minimize the variance to the Poisson level and account for the survey geometry.

Each galaxy carries several weights to correct for possible observational effects; therefore, the number counts for galaxy pairs should be weighted as follows :
\begin{align}\label{eq:weight}
GG(r_\alpha)&=\sum\limits_{G_i,G_j}w(G_i)w(G_j)\Theta_{r_\alpha}(d_{ij}),\,G=D,R;\\
\Theta_{r_\alpha}(d_{ij})&=\left\{
                         \begin{array}{ll}
                           1, & \hbox{$r_\alpha\leq d_{ij}<r_\alpha+\Delta r$;} \\
                           0, & \hbox{otherwise,}
                         \end{array}
                       \right.
\end{align}
where $d_{ij}$ is the separation between  pairwise galaxies $G_i$ and $G_j$, and $G_{i,j}$ stands for galaxy $i,j$. The specific choices of the weights for the data galaxy $w(D)$ and random galaxy $w(R)$ will be presented in Section \ref{sec:methodology}.

\subsection{Fitting Template}\label{subsec:fitting}

The fitting model~\citep{Anderson:2012sa,Anderson:2013oza,Xu:2012fw} for the monopole moment of 2PCF is parametrized as
\begin{equation}\label{eq:2PCF}
\xi_\mathrm{model}(r)=B_0^2\xi_\mathrm{theory}(\alpha r)+A_0+\frac{A_1}{r}+\frac{A_2}{r^2},
\end{equation}
where the first nuisance parameter $B_0$ is a normalization factor, and the other three nuisance parameters $A_{0,1,2}$ are introduced to account for any broad-band derivations from the theoretical 2PCF $\xi_\mathrm{theory}(r)$. The dilation parameter $\alpha$ is defined by
\begin{equation}\label{eq:alpha}
\alpha=\frac{D_V(z)/r_\mathrm{d}}{\left(D_V(z)/r_\mathrm{d}\right)_\mathrm{fiducial}}.
\end{equation}
The BAO peak position shifts towards a smaller (larger) scale if $\alpha>1$ ($\alpha<1$) with respect to the fiducial cosmology.

The $l$-th moment of theoretical 2PCF $\xi_\mathrm{theory}^{(l)}(r)$ is the Fourier transformation
\begin{equation}
\xi_\mathrm{theory}^{(l)}(r)=i^l\int\frac{k^2\mathrm{d}k}{2\pi^2}j_l(kr)P_\mathrm{nonlinear}^{(l)}(k)
\end{equation}
of the $l$-th multipole of the non-linear power spectrum $P_\mathrm{nonlinear}^{(l)}(k)$,
\begin{equation}
P_\mathrm{nonlinear}^{(l)}(k)=\frac{2l+1}{2}\int_{-1}^1\mathrm{d}\mu L_l(\mu)P_\mathrm{nonlinear}(k,\mu),
\end{equation}
The anisotropic non-linear power spectrum $P_\mathrm{nonlinear}(k,\mu)$~\citep{Fisher:1993pz} is modelled as
\begin{equation}
P_\mathrm{nonlinear}(k,\mu)=\frac{(1+\beta\mu^2)^2}{(1+k^2\mu^2\Sigma_s^2)^2}P_\mathrm{dewiggle}(k,\mu),
\end{equation}
where $(1+\beta\mu^2)^2$ accounts for the Kaiser effect~\citep{Kaiser:1987qv} and $1/(1+k^2\mu^2\Sigma_s^2)^2$ for the Finger-of-God (FoG) effect~\citep{Park:1994fa}. $\beta=f/b$ where $f\approx\Omega_\mathrm{m}(z)^{0.55}$ is the growth rate and $b$ is the galaxy bias with respect to dark matter. The de-wiggled power spectrum~\citep{Eisenstein:2006nj,Anderson:2012sa,Anderson:2013oza,Xu:2012fw,Vargas-Magana:2015rqa} models the degradation of the BAO feature due to non-linear structure growth :
\begin{align}
P&_\mathrm{dewiggle}(k,\mu)=\left[P_\mathrm{linear}(k)-P_\mathrm{nowiggle}(k)\right]\nonumber\\
&\times\exp\left[-\frac{k^2\mu^2\Sigma_\parallel^2+k^2(1-\mu^2)\Sigma_\perp^2}{2}\right]+P_\mathrm{nowiggle}(k).
\end{align}
where both the linear theory power spectrum $P_\mathrm{linear}(k)$~\citep{Lewis:1999bs} and no-wiggle power spectrum $P_\mathrm{nowiggle}(k)$~\citep{Eisenstein:1997ik} are well approximated in details in~\cite{Eisenstein:1997ik}.

We focus only on the monopole of 2PCF with the pre-reconstructed samples. The anisotropic fitting of a higher moment of 2PCF with reconstructed samples will be reserved for future works. In this study, we set the streaming scale $\Sigma_s=4h^{-1}\,\mathrm{Mpc}$, the radial component $\Sigma_\parallel=10h^{-1}\,\mathrm{Mpc}$ and transverse component $\Sigma_\perp=6h^{-1}\,\mathrm{Mpc}$ of the standard non-linear Gaussian damping scale $\Sigma_\mathrm{non-linear}^2=(\Sigma_\parallel^2+\Sigma_\perp^2)/2$ as inferred from~\cite{Vargas-Magana:2015rqa}.
The spatial separations of interest are chosen from $4h^{-1}\,\mathrm{Mpc}$ to $204h^{-1}\,\mathrm{Mpc}$ with a binning size of $8h^{-1}\,\mathrm{Mpc}$. The fitting range is between $30\sim180 h^{-1}\,\mathrm{Mpc}$ also with a binning size of $8h^{-1}\,\mathrm{Mpc}$ as in the SDSS-III BOSS DR12~\citep{Cuesta:2015mqa}. Therefore, the total number of points to be fitted is 19 and the fitting parameters to be used are $B_0, \alpha, \beta, A_0, A_1, A_2$, leaving $13$ degrees of freedom. The initial values and Gaussian priors $[0\pm0.4,1\pm0.1,0.4\pm0.2]$ are adopted for $[\log B_0^2, \alpha, \beta]$.

The $\chi^2$ goodness-of-fit indicator is defined by
\begin{equation}\label{eq:chisq}
\chi^2=\sum\limits_{i,j}[\xi_\mathrm{model}(r_i)-\xi_\mathrm{data}(r_i)]C_{ij}^{-1}[\xi_\mathrm{model}(r_j)-\xi_\mathrm{data}(r_j)],
\end{equation}
where the covariance matrix is given by
\begin{equation}\label{eq:covariance}
C_{ij}=\frac{1}{N_{\mathrm{mock}}-1}\sum\limits_{n=1}^{N_\mathrm{mock}}[\xi_n(r_i)-\bar{\xi}_\mathrm{mock}(r_i)][\xi_n(r_j)-\bar{\xi}_\mathrm{mock}(r_j)],
\end{equation}
Here $\bar{\xi}_\mathrm{mock}(r_i)$ is the mean value averaging over the corresponding 2PCF from all QPM mock catalogues at separation bin $r_i$. To have a unbiased estimation of the inverse covariance matrix~\citep{Hartlap:2006kj}, one has to multiply the covariance matrix \eqref{eq:covariance} by
\begin{align}
C_{ij}\rightarrow\frac{N_{\mathrm{mock}}-1}{N_{\mathrm{mock}}-N_{\mathrm{data}}-2}C_{ij},
\end{align}
where the number of mocks is $N_{\mathrm{mock}}=1000$ and the size of data vector is $N_{\mathrm{data}}=19$. One can also marginalize over the other five parameters for the fixed value of $\alpha$ and obtain the probability density function (PDF) as
\begin{align}
p(\alpha_i)=\exp\left(-\frac{\chi^2(\alpha_i)}{2}\right)\left/\int\mathrm{d}\alpha_i\exp\left(-\frac{\chi^2(\alpha_i)}{2}\right)\right.,
\end{align}
based on which we can calculate the mean value $\langle\alpha\rangle$ and the variance $\sigma_\alpha^2=\langle\alpha^2\rangle-\langle\alpha\rangle^2$.

\section{Random Catalogues}\label{sec:methodology}

In this section, we describe the wiggly scheme and smooth scheme of redshift distributions used to generate the random galaxy catalogues, which are summarized in Table \ref{tab:nz}.
\begin{table*}
	\centering
	\caption{Galaxy sample and random sample with a wiggly method and a smooth method both for the DR12 data and for the QPM mocks. $N_\mathrm{g}(z)$ is the observed redshift distribution and $N_\mathrm{r}$ is the redshift distribution generated with a wiggle method. $N_{\mathrm{g},i}$ is the redshift distribution of the $i$-th mock sample, $N_\mathrm{s}$ is the observed redshift distribution weighted by $w_\mathrm{sys}$, $N_w$ is the observed redshift distribution weighted by $w_\mathrm{sys}$ and $w_\mathrm{FKP}$, and $N_{\mathrm{w},i}$ are the redshift distribution of the $i$-th mock sample weighted by $w_\mathrm{sys}$ and $w_\mathrm{FKP}$. Here $w_\mathrm{sys}$ is the systematic weight accounting for the observational systematic effects, and $w_\mathrm{FKP}$ is the FKP weight to reduce the Poisson noise.}
	\label{tab:nz}
    \renewcommand\arraystretch{1.2}
	\begin{tabular}{c|c|c|c|c|c|c|c|c}
     \hline
     \hline
     \multirow{4}*{} & \multicolumn{4}{c|}{Wiggly method}                             & \multicolumn{4}{c}{Smooth method}\\
                       \cline{2-5}\cline{6-9}
                     & \multicolumn{2}{c|}{DR12 data} & \multicolumn{2}{c|}{QPM mock} & \multicolumn{2}{c|}{DR12 data} & \multicolumn{2}{c}{QPM mock}\\
                       \cline{2-5}\cline{6-9}
                     & Redshift       & Weight for   & Redshift        & Weight for   & Redshift        & Weight for   & Redshift        & Weight for \\
                     & population     & pair-count   & population      & pair-count   & population      & pair-count   & population      & pair-count \\
     \hline
	 Galaxy          & \multicolumn{1}{c|}{\multirow{2}*{$N_\mathrm{g}(z)$}}
                     & \multicolumn{1}{c|}{\multirow{2}*{$w_{\mathrm{sys}}\cdot w_{\mathrm{FKP}}$}}
                     & \multicolumn{1}{c|}{\multirow{2}*{$\left\{N_{\mathrm{g},i}(z)\right\}_{i=1}^{1000}$}}
                     & \multicolumn{1}{c|}{\multirow{2}*{$w_{\mathrm{sys}}\cdot w_{\mathrm{FKP}}$}}
                     & \multicolumn{1}{c|}{\multirow{2}*{$N_\mathrm{g}(z)$}}
                     & \multicolumn{1}{c|}{\multirow{2}*{$w_{\mathrm{sys}}\cdot w_{\mathrm{FKP}}$}}
                     & \multicolumn{1}{c|}{\multirow{2}*{$\left\{N_{\mathrm{g},i}(z)\right\}_{i=1}^{1000}$}}
                     & \multicolumn{1}{c}{\multirow{2}*{$w_{\mathrm{sys}}\cdot w_{\mathrm{FKP}}$}}\\
     sample          & & & & & & & & \\
     \hline
     Random          & \multicolumn{1}{c|}{\multirow{2}*{$\frac{N_\mathrm{r}(z)}{N_\mathrm{s}(z)}\approx50$}}
                     & \multicolumn{1}{c|}{\multirow{2}*{$w_{\mathrm{FKP}}$}}
                     & \multicolumn{1}{c|}{\multirow{2}*{$\frac{N_\mathrm{r}(z)}{N_{\mathrm{s},i}(z)}\approx50$}}
                     & \multicolumn{1}{c|}{\multirow{2}*{$w_{\mathrm{FKP}}$}}
                     & \multicolumn{1}{c|}{\multirow{2}*{$\begin{array}{c}\mathrm{fitting}\\N_\mathrm{w}(z)\end{array}$}}
                     & \multicolumn{1}{c|}{\multirow{2}*{$1$}}
                     & \multicolumn{1}{c|}{\multirow{2}*{$\big\{\begin{array}{c}\mathrm{fitting}\\N_{\mathrm{w},i}(z)\end{array}\big\}_{i=1}^{1000}$}}
                     & \multicolumn{1}{c}{\multirow{2}*{$1$}}\\
     sample          & & & & & & & & \\
     \hline
     \hline
	\end{tabular}
\end{table*}

\subsection{Wiggly Method}\label{subsec:wigglymethod}

The recent analysis~\citep[e.g.][]{Reid:2015gra} of galaxy clustering utilizes the observed redshift distribution of data galaxy catalogues to generate the random galaxy catalogues. This can be mimicked using the following procedure. With a redshift bin size $\Delta z=0.001$, the observed data galaxy redshift distribution for each redshift bin $z_\alpha$,
\begin{align}
\begin{split}\label{eq:ng}
N_\mathrm{g}(z_\alpha)&=\sum\limits_{D_i}\Theta_{z_\alpha}(z_{D_i}),\\
\Theta_{z_\alpha}(z_{D_i})&=\left\{
                         \begin{array}{ll}
                           1, & \hbox{$z_\alpha\leq z_{D_i}<z_\alpha+\Delta z$;} \\
                           0, & \hbox{otherwise,}
                         \end{array}
                       \right.
\end{split}
\end{align}
is a wiggly curve along the redshift direction, of which the fluctuations could contain both BAO signals and spurious fluctuations that are caused by observational systematic effects and Poisson noises.

The observational systematic effects can be corrected by assigning each data galaxy with a systematic weight~\citep{Reid:2015gra} given by
\begin{equation}\label{eq:wsys}
w_\mathrm{sys}(D_i)=w_{\mathrm{star},i}w_{\mathrm{see},i}(w_{\mathrm{cp},i}+w_{\mathrm{noz},i}-1),
\end{equation}
where $w_{\mathrm{star},i}w_{\mathrm{see},i}$ are the total angular systematic weights that contribute little to those wiggles of redshift distribution $N_\mathrm{g}(z)$. The weights $w_{\mathrm{cp},i}$ that correct for close pairs (fibre collisions) and the weights $w_{\mathrm{noz},i}$ that correct for redshift failures have a little impact on the measured clustering. Therefore, the redshift distribution $N_\mathrm{s}(z)$ corrected by these systematic weights is given by
\begin{align}
\label{eq:ns}
N_\mathrm{s}(z_\alpha)&=\sum\limits_{D_i}w_\mathrm{sys}(D_i)\Theta_{z_\alpha}(z_{D_i}),
\end{align}
The wiggly method thus generates random catalogues with a redshift distribution
\begin{equation}\label{eq:nr}
N_\mathrm{r}(z)\sim N_\mathrm{s}(z).
\end{equation}
Note that the total number of random galaxies is usually larger than the total number of data galaxies, $N_\mathrm{r}(z)=50\times N_\mathrm{s}(z)$, for example in SDSS-III BOSS DR12~\citep{Reid:2015gra}.

The theoretical Poisson noises can also be corrected during pair-counting by assigning each galaxy of pairs the Feldman-Kaiser-Peacock (FKP) weight~\citep{Feldman:1993ky} given by
\begin{equation}
w_{\mathrm{FKP},i}=\frac{1}{1+n(z_i)P_0},
\end{equation}
where $n(z_i)$ is the measured number \emph{density} at redshift $z_i$ with linear interpolation over bins $\Delta=0.005$ and $P_0=20000h^{-3}\,\mathrm{Mpc}^3$ is the observed power spectrum at $k\approx0.15h\,\mathrm{Mpc}^{-1}$. The wiggly method assigns total weights to each galaxy of pairs in different ways depending on its type,
\begin{align}
\begin{split}\label{eq:wigglypaircounting}
GG(r_\alpha)&=\sum\limits_{G_i,G_j}w(G_i)w(G_j)\Theta_{r_\alpha}(d_{ij});\\
w(D_i)&=w_{\mathrm{sys},i}\cdot w_{\mathrm{FKP},i};\\
w(R_i)&=w_{\mathrm{FKP},i},
\end{split}
\end{align}
where pairwise galaxies $GG$ can be of $DD$, $RR$ and $DR$ type.

The wiggly method also applies to each QPM mock catalogue labelled by $i$ from 1 to 1000, with their mock random catalogue having the same redshift distribution \eqref{eq:nr} as the systematically weighted redshift distribution of the mock galaxy catalogue,
\begin{align}\label{eq:mocknr}
N_{\mathrm{r},i}(z)\sim N_{\mathrm{s},i}(z).
\end{align}
The weights assignments for QPM mock galaxy/random catalogues during pair-counting are the same as \eqref{eq:wigglypaircounting} as we do for the DR12 data galaxy/random catalogues.

Note in the QPM mock catalogue, each galaxy is assigned with the FKP weight and `veto mask', where the `veto mask' is $w_{\mathrm{cp},i}+w_{\mathrm{noz},i}-1$. No systematic weights are assigned to mock galaxies in the sense to correct for systematics in the imaging data, i.e. $w_{\mathrm{star},i}w_{\mathrm{see},i} = 1$. We keep the same notation for simplicity.

\subsection{Smooth Method}\label{subsec:smoothmethod}

Besides systematic effects and Poisson noises, the observed reshift distribution also could contains fluctuations caused by real structures. A true random catalogue should remove all the spurious fluctuations and only retain a smooth shape of redshift distribution. This motivates us to propose the following smooth method.

When assigning redshifts to the random catalogues, all weights should be taken into account. The observed redshift distribution of data galaxy catalogues, after weighted by both systematic weights and FKP weights,
\begin{align}
\label{eq:nw}
N_\mathrm{w}(z_\alpha)&=\sum\limits_{D_i}w_\mathrm{sys}(D_i)w_\mathrm{FKP}(D_i)\Theta_{z_\alpha}(z_{D_i}),
\end{align}
gives rise to a less wiggly redshift distribution $N_\mathrm{w}(z)$, which can be fitted with some smooth functions $N_\mathrm{f}(z)$. Here we adopt the Moffat function,
\begin{equation}
N_f(z)=\frac{a_0}{\left(\left(\frac{z-a_1}{a_2}\right)^2+1\right)^{a_3}}+a_4+a_5z.
\end{equation}
Therefore the smooth method generates random catalogues with a redshift distribution,
\begin{equation}\label{eq:nf}
N_\mathrm{r}(z)\sim N_\mathrm{f}(z),
\end{equation}
and assigns weight for each galaxy of galaxy pairs by
\begin{align}
\begin{split}\label{eq:smoothpaircounting}
GG(r_\alpha)&=\sum\limits_{G_i,G_j}w(G_i)w(G_j)\Theta_{r_\alpha}(d_{ij});\\
w(D_i)&=w_{\mathrm{sys},i}\cdot w_{\mathrm{FKP},i};\\
w(R_i)&=1
\end{split}
\end{align}
during pair-counting.

The same strategies also apply to each QPM mock catalogue, where its random sample is generated with redshift distribution given by
\begin{equation}\label{eq:mocknf}
N_{\mathrm{r},i}\sim N_{\mathrm{f},i}(z),
\end{equation}
The weights assignments for QPM mock galaxy/random catalogues for pair-counting are the same \eqref{eq:smoothpaircounting} as what we do for the DR12 data galaxy/random catalogues.

\begin{figure*}
	\includegraphics[width=\textwidth]{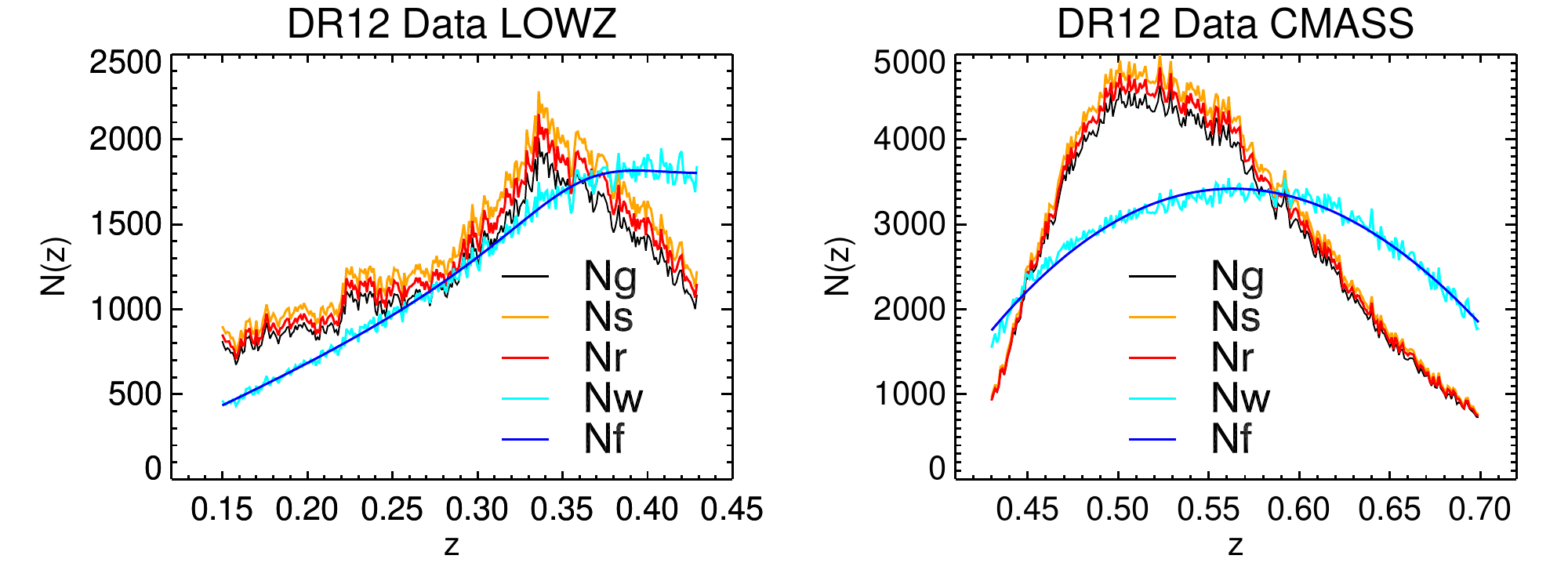}
    \caption{Redshift distributions of the DR12 data catalogs for LOWZ sample (left-hand panel) and CMASS sample (right-right panel). Different samples are presented with different colours as indicated in the each panel. See the text and Table \ref{tab:nz} for the definition of each sample. It shows that the smooth method eliminates those wiggles presented in $N_\mathrm{w}(z)$. All redshift distributions have been scaled properly for clearness.}
    \label{fig:nz}
\end{figure*}

As a comparison, we present various redshift distributions in Fig.\ref{fig:nz}
for SDSS-III BOSS DR12 data catalogues. As we can see, the random catalogues generated by wiggly method in line with the redshift distribution from the systematically weighted redshift distribution, while the smooth method eliminates the wiggles that are caused by structures.

\subsection{Tests on Mock Samples}\label{subsec:mock}

\begin{figure*}
	\includegraphics[width=\textwidth]{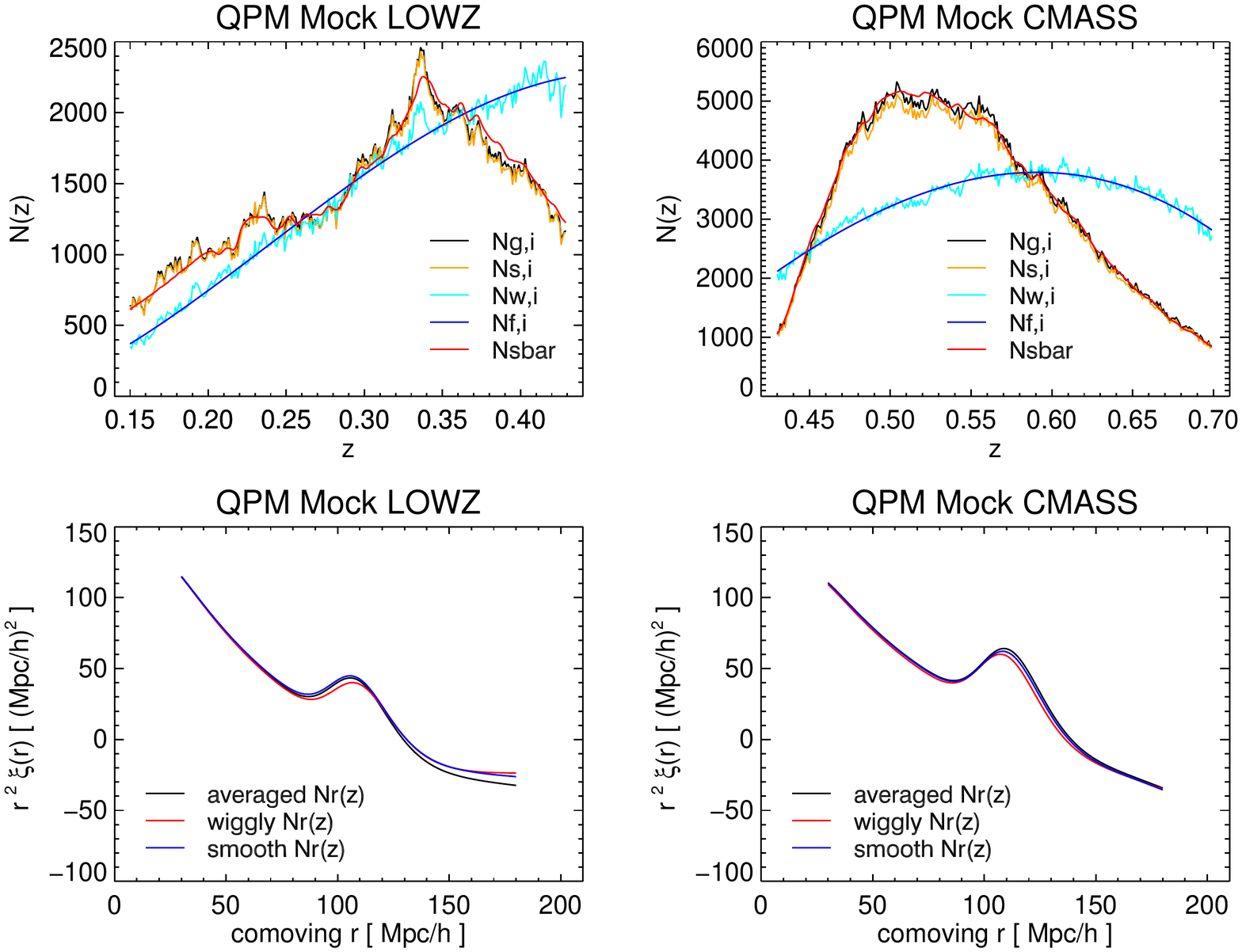}
    \caption{Top panels: redshift distribution for one particular QPM mock catalogue (the $i$-th one) for both LOWZ (left-hand column) and CMASS (right-hand column) samples. Different colours are for different sample definitions as denoted in the right-hand corner of each panel. The fluctuation of $N_{\mathrm{w},i}$ is rather big to illustrate the effect. Bottom panels: 2PCFs calculated with random samples generated with the wiggly method, smooth method and the true redshift distribution are shown with red, blue and black curves, respectively. Results with the smooth method are closer to the true ones compared to the wiggly method.}
    \label{fig:mocktest}
\end{figure*}

Before applying to the DR12 data catalogues, we test our smooth method against the wiggly method on QPM mock catalogues. We first select one particular mock that has a rather large scatter in $N_g(z)$ and can thus reflects more clearly the difference caused by different random samples. Three random samples are generated with the wiggly reshift distribution $N_\mathrm{s}(z)$, the smooth redshift distribution $N_\mathrm{f}(z)$ and an underlying true redshift distribution, respectively. Here the true redshift distribution is assumed to be the averaged value of systematically weighted redshift distribution over 1000 QPM mock galaxy samples,
\begin{equation}\label{eq:nsbar}
\overline{N}_\mathrm{s}(z)=\frac{1}{N_{\mathrm{mock}}}\sum_{i=1}^{N_{\mathrm{mock}}} N_{\mathrm{s},i}(z),
\end{equation}
which is the prior redshift distribution used for generating those mock catalogues.
The weight assignment during pair-counting for the random catalogue generated by \eqref{eq:nsbar} is as same as the wiggly method \eqref{eq:wigglypaircounting}.
We will refer to the 2PCF calculated from above mock random catalogue with redshift distribution $\overline{N}_\mathrm{s}(z)$ as the true 2PCF in following.
As shown in the bottom panels of Fig.\ref{fig:mocktest},
the measurements of the 2PCF at BAO scale are closer to the true one with our smooth method than the results with the wiggly method.
The dilation parameters along with their $1\sigma$ errors given by the wiggly method, smooth method and true redshift distribution are $0.9835\pm0.0321$, $0.9949\pm0.0305$ and $0.9932\pm0.0295$, respectively, for LOWZ sample, and $0.9840\pm0.0161$, $0.9746\pm0.0152$ and $0.9711\pm0.0154$, respectively, for CMASS sample. The volume-averaged distances $D_V(z)r_{\mathrm{d}}^{\mathrm{fid}}/r_{\mathrm{d}}$ are $1215\pm40$, $1229\pm38$ and $1227\pm36$, respectively, for LOWZ sample, and $1977\pm32$, $1958\pm30$ and $1952\pm31$, respectively, for CMASS sample. It shows that both the dilation parameter and the volume-averaged distance are close to the true ones with the smooth method. The $1\sigma$ errors are also reduced when adopting the smooth method.

We further apply this procedure to the 1000 mock samples to investigate whether our smooth method can also boost the BAO signals and reduce the error of dilation parameter in general.

\begin{figure*}
  \includegraphics[width=\textwidth]{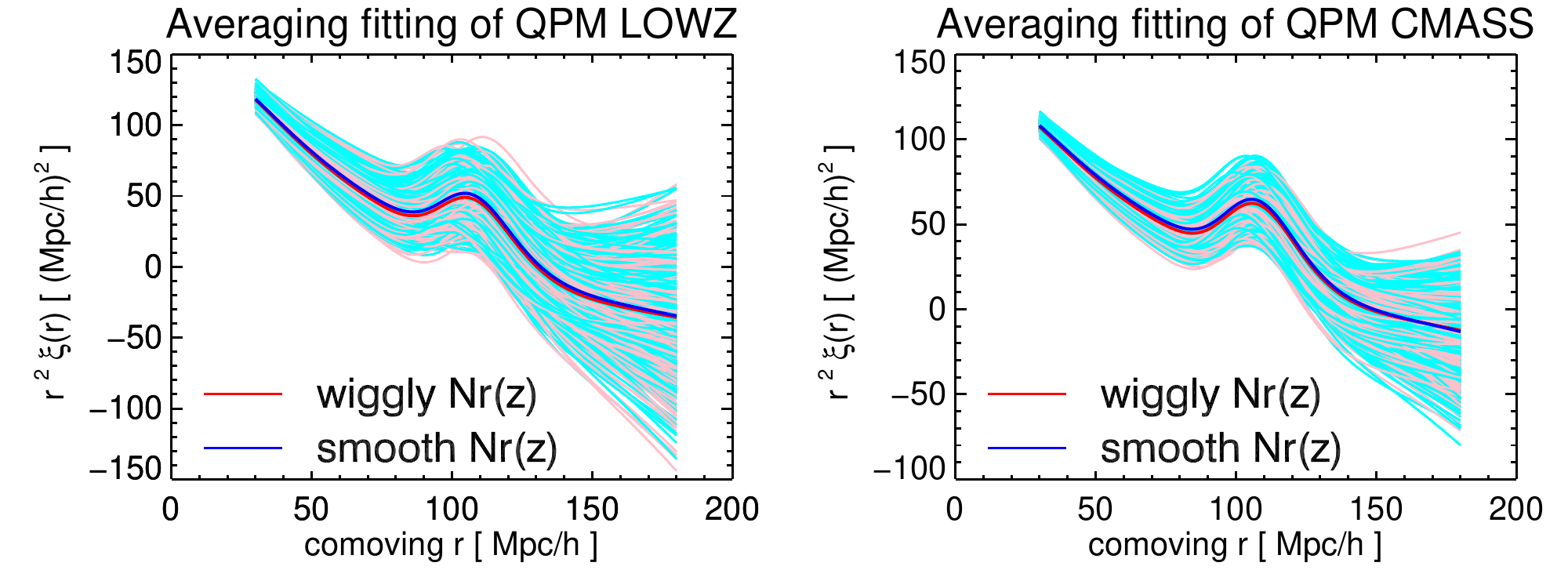}
  \caption{The fitting results of 2PCF of all the 1000 LOWZ (left-hand panel) and CMASS (right-hand panel) mock samples with the wiggly method (red) and the smooth method (blue). Each thin curve represents for one mock results. The averaged 2PCFs are presented with thick curves. The 2PCF at BAO scale is slightly lifted with the smooth method over the wiggly method both for LOWZ and CMASS mock samples.}
  \label{fig:mockavrg}
\end{figure*}

Fitting results for the 1000 mocks are shown in Fig.\ref{fig:mockavrg}
with thin curves. Averaged fits are shown with thick curves. As we can see, there is a gentle lift of BAO signals with the smooth method compared to the wiggly method. As in~\cite{Cuesta:2015mqa}, we can use this 1000 mock samples to study the distribution of the dilation parameter in order to quantify the difference of these two methods. We found that, with wiggly method, the mean value of dilation parameter $\langle\alpha\rangle$ and its $1\sigma$ error $\langle\sigma_\alpha\rangle$ and the standard derivation $S_\alpha$ are $1.00183$, $0.02994$ and $0.02856$ for LOWZ sample and $1.00035$, $0.01834$ and $0.01715$ for CMASS sample, respectively. With the smooth method, the mean value of dilation parameter $\langle\alpha\rangle$ and its $1\sigma$ error $\langle\sigma_\alpha\rangle$ and the standard derivation $S_\alpha$ are $1.00250$, $0.02931$ and $0.02794$ for LOWZ sample and $1.00084$, $0.01766$ and $0.01664$ for CMASS sample, respectively. It is obvious that, with our smooth method, the mean value of $1\sigma$ error $\langle\sigma_\alpha\rangle$ and the standard derivation $S_\alpha$ are reduced both for the LOWZ and the CMASS mock samples, though the magnitude is only of the order of $\mathcal{O}(0.001)$, similar to other systematic effects~\citep{Vargas-Magana:2016imr}.

Note that in the literatures, a shuffled method \citep[e.g.][]{Ross:2012qm,Reid:2015gra} is usually adopted to assign redshifts to the random galaxies by randomly drawing from the measured galaxy redshifts. \cite{Ross:2012qm} demonstrated that a $N$-node spline method could reproduce the 2PCF estimated with the shuffled method, when $N$ is large. The wiggly method is very close to the $N$-node spline method and we adopted a very small redshift interval, $\Delta z = 0.001$, corresponding to a very large $N$. The resulting 2PCF estimated using the wiggly method is thus very close to the one estimated using the shuffled method. As shown above, the smooth method behave better than the wiggly method. We thus could conclude that it  also works better than the shuffled method adopted in the literatures.

\section{Results}\label{sec:data}

\begin{figure*}
	\includegraphics[width=\textwidth]{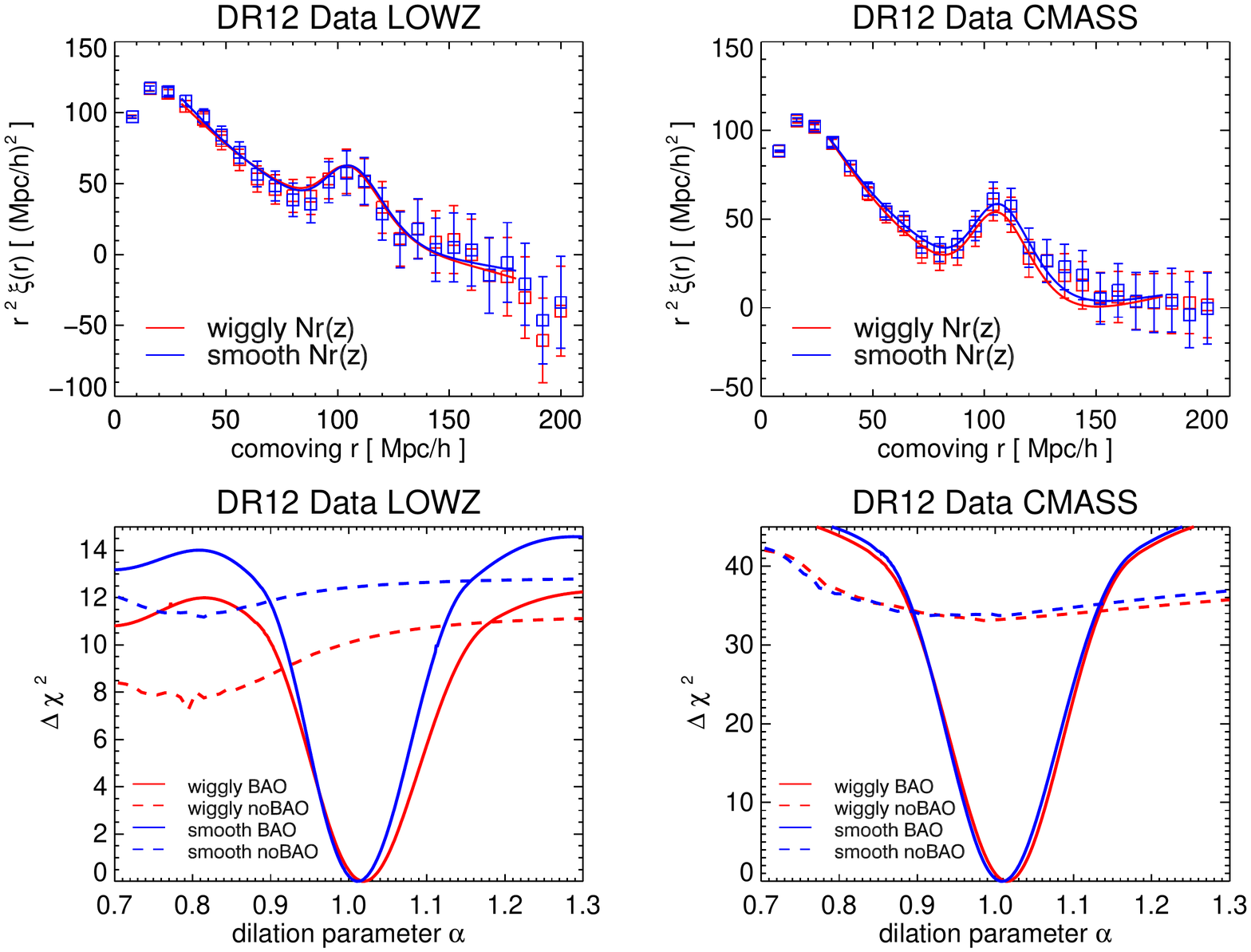}
    \caption{The 2PCF (top) and $\chi^2$ surface (bottom) for DR12 data catalogues are shown for LOWZ sample (left) and CMASS sample (right) with the wiggly method (red) and the smooth method (blue). In the top panels, there is a gentle lift of BAO signals with the smooth method over the wiggly method. In the bottom panels, it shows a minor improvement of detection significance for the BAO signals.}
    \label{fig:datamore}
\end{figure*}

In this section, we apply our smooth method to the SDSS-III BOSS DR12 data catalogues for both LOWZ and CMASS samples and compare the results to those with the wiggly method. The results are presented in Fig.\ref{fig:datamore}
and Table \ref{tab:data}.
\begin{table}
	\centering
	\caption{Results for the DR12 data catalogues. In the upper part of the table, the best-fitting results of dilation parameter $\alpha$ are shown along with their $1\sigma$ errors $\sigma_\alpha$, which can be used to constrain the volume-average distance to the effective redshifts of LOWZ sample $z=0.32$ and CMASS sample $z=0.57$. In the lower part of the table, the mean value of dilation parameter $\alpha$ are shown along with their $1\sigma$ errors $\sigma_\alpha=\sqrt{\langle\alpha^2\rangle-\langle\alpha\rangle^2}$. The detection significance is calculated from the square root of difference of $\chi^2$ from the de-wiggled and no-wiggled template at the minimum of $\chi^2$ surface. The upshot of these results is that there is a gentle but uniform improvement on the error of dilation parameter and the detection significance for the DR12 data catalogues.}
	\label{tab:data}
    \renewcommand\arraystretch{1.5}
	\begin{tabular}{cccc}
     \hline
     \hline
     2PCF $\xi(r)$ & $\alpha\pm\sigma_\alpha$ & $D_V(z)\frac{r_{\mathrm{d}}^{\mathrm{fid}}}{r_{\mathrm{d}}}$ & $\chi^2/\mathrm{dof}$ \\
     \hline
     \hline
     DR12 LOWZ     &                   &              &           \\
     Wiggly method & $1.0175\pm0.0298$ & $1257\pm37$  & $8.5/13$  \\
     Smooth method & $1.0105\pm0.0288$ & $1248\pm36$  & $8.5/13$  \\
     \hline
     DR12 CMASS    &                   &              &           \\
     Wiggly method & $1.0147\pm0.0163$ & $2039\pm33$  & $14/13$   \\
     Smooth method & $1.0076\pm0.0158$ & $2025\pm32$  & $14/13$   \\
     \hline
     \hline
     PDF $p(\alpha)$ & $\langle\alpha\rangle$ & $\sigma_\alpha$ & $\sqrt{\Delta\chi_{\mathrm{min}}^2}$ \\
     \hline
     \hline
     DR12 LOWZ       &                        &             &               \\
     Wiggly method   & $1.0202$               & $0.0422$    & $3.20\sigma$  \\
     Smooth method   & $1.0126$               & $0.0323$    & $3.53\sigma$  \\
     \hline
     DR12 CMASS      &                        &             &               \\
     Wiggly method   & $1.0148$               & $0.0157$    & $5.77\sigma$  \\
     Smooth method   & $1.0087$               & $0.0154$    & $5.80\sigma$  \\
     \hline
     \hline
	\end{tabular}
\end{table}

In the top panels of Fig.\ref{fig:datamore}, the 2PCFs are shown for the LOWZ and CMASS samples with the wiggly method and the smooth method. Red curves are the best-fitting results and error bars are obtained from the square root of the diagonal elements of the covariance matrix, which is obtained from the QPM mock catalogues. As we can see, there is a gentle lift of BAO signals with our smooth method with respect to the wiggly method. This exactly meets our expectation for the smooth method, where a more randomly populated catalogue along the redshift direction necessarily wipes out residual fluctuations of clustering signals inherited from the measured redshift distribution. Therefore a more pronounced BAO signal is obtained with the smooth method.

In the bottom panels of Fig.\ref{fig:datamore}, we present the likelihood surfaces for a grid of the fixed dilation parameter $\alpha$ from isotropically fitting the 2PCF for the LOWZ sample (left-hand panel) and the CMASS sample (right-hand panel) with the wiggly method (red) and the smooth method (blue) using fitting templates with (solid) and without (dashed) the BAO feature. Each likelihood surface has been subtracted with their minimum $\chi^2$ values. The difference in $\chi^2$ between the template with and without BAO feature reflects the significance of detection of BAO. The $\chi^2$ minimum is slightly narrowed down with our smooth method compared to the wiggly method, which indicates a minor improvement on the constraint of dilation parameter. The upshot of the bottom panels of Fig.\ref{fig:datamore} is that both the significance of detection of BAO and the constraint on the dilation parameter have been improved gently with our smooth method over the wiggly method.

In Table \ref{tab:data}, we briefly summarize the fitting results of 2PCF for LOWZ sample and CMASS sample. In the upper part of the Table \ref{tab:data}, the best-fitting results of dilation parameter $\alpha$ are shown along with their $1\sigma$ errors $\sigma_\alpha$, which can be used to constrain the volume-average distances to the effective redshifts of LOWZ sample $z=0.32$ and CMASS sample $z=0.57$. Although the shift of dilation parameter is minor between the smooth method and wiggly method, both LOWZ and CMASS samples have exhibited an uniform reduction on the error of dilation parameter with our smooth method over the wiggly method. In the lower part of the Table \ref{tab:data}, we present the fitting results from probability density function (PDF) of  for LOWZ sample and CMASS sample. The mean values of dilation parameter $\alpha$ are shown along with their $1\sigma$ errors $\sigma_\alpha=\sqrt{\langle\alpha^2\rangle-\langle\alpha\rangle^2}$ and the detection significances. The significance of detection of BAO has been improved by $0.33\sigma$ for LOWZ sample and $0.03\sigma$ for CMASS sample. Although the boosted signals and reduced error are necessarily small, it would be appreciated for the future improvements on the measurements of galaxy clustering given the high precise of the measurements of cosmological parameters.

\section{Conclusions}\label{sec:conclusions}

The measurements of galaxy clustering on BAO scale have reached an unprecedented precision since the DR11 of BOSS of SDSS-III. Further improvements require more careful understanding of the potential errors. The 2PCF of galaxy clustering measures the number excesses of the data-data galaxy pairs with respect to the random-random galaxy pairs. Therefore, a random sample with proper redshift distributions is necessary to recover the BAO signals. In this paper, we proposed to use a smooth function that fits the observed galaxy redshift distribution to generate the random galaxy catalogue. It has the advantage to remove the BAO signals along the redshift direction compared with the wiggly method that is usually adopted in the literatures, which assigns each random galaxy a redshift by randomly drawing a value from the observed redshift distribution, and can thus boost the measured BAO signals. Using the mock data, we demonstrated that the smooth method is capable of improving BAO signals and reducing the errors compared with the wiggly method. When applying to the SDSS DR12 data, we found with the smooth method, we can improve the significance of BAO signals by 0.33 and 0.03$\sigma$ for the LOWZ and CMASS samples, respectively. The rather small improvements is expected, because both the LOWZ and CMASS samples cover rather large volume and the sample size is large, which means that the observed redshift distribution suffers little from cosmic variance. For studies that contain smaller samples, for example, the QSO sample, a smooth method could be more helpful in improving the results. Moreover, as discussed in~\cite{Ross:2012qm}, although the effect of redshift distribution on the monopole of the clustering could be ignored, its effect is large for multipoles. We will explore this effect with post-reconstruction data in future works.

\section*{Acknowledgements}

We want to thank Antonio J. Cuesta, Benjamin Alan Weaver, and Jeremy Tinker of SDSS-III BOSS DR12 for the help on BAO fitting and weight assignment of QPM mocks. We also want to thank Yuting Wang, Gong-Bo Zhao and Bin Hu for helpful discussions. SJW wants to thank Shuangpeng Sun, Shihong Liao, Junyi Jia, Zhen Jiang and Chunxiang Wang for the help on \textsc{IDL} coding, and Wei-Ming Dai and Li-Wei Ji for the help on \textsc{Fortran} coding during this work. S-JW also wants to thank the hospitality of Misao Sasaki when the work was finished during the visit at the Center for Gravitational Physics of Yukawa Institute for Theoretical Physics of Kyoto University. We acknowledge the use of the cluster PANGU and NOVA of NAOC.
QG is supported by a Newton Advanced Fellowship, NSFC grant (No. 11573033, 11622325), the Strategic Priority Research Program ``The Emergence of Cosmological Structure'' of the Chinese Academy of Sciences (No.XDB09000000) and the ``Recruitment Program of Global Youth Experts'' of China, the NAOC grant (Y434011V01).
R-GC is supported in part by the Strategic Priority Research Program of the Chinese Academy of Science (CAS), Grant No.XDB09000000 and by a key project of CAS, Grant No.QYZDJ-SSW-SYS006 and by the National Natural Science Foundation of China under Grants No.11375247 and No.11435006.

\bibliographystyle{mnras}
\bibliography{ref}

\begin{thebibliography}{}
\makeatletter
\relax
\def\mn@urlcharsother{\let\do\@makeother \do\$\do\&\do\#\do\^\do\_\do\%\do\~}
\def\mn@doi{\begingroup\mn@urlcharsother \@ifnextchar [ {\mn@doi@}
  {\mn@doi@[]}}
\def\mn@doi@[#1]#2{\def\@tempa{#1}\ifx\@tempa\@empty \href
  {http://dx.doi.org/#2} {doi:#2}\else \href {http://dx.doi.org/#2} {#1}\fi
  \endgroup}
\def\mn@eprint#1#2{\mn@eprint@#1:#2::\@nil}
\def\mn@eprint@arXiv#1{\href {http://arxiv.org/abs/#1} {{\tt arXiv:#1}}}
\def\mn@eprint@dblp#1{\href {http://dblp.uni-trier.de/rec/bibtex/#1.xml}
  {dblp:#1}}
\def\mn@eprint@#1:#2:#3:#4\@nil{\def\@tempa {#1}\def\@tempb {#2}\def\@tempc
  {#3}\ifx \@tempc \@empty \let \@tempc \@tempb \let \@tempb \@tempa \fi \ifx
  \@tempb \@empty \def\@tempb {arXiv}\fi \@ifundefined
  {mn@eprint@\@tempb}{\@tempb:\@tempc}{\expandafter \expandafter \csname
  mn@eprint@\@tempb\endcsname \expandafter{\@tempc}}}

\bibitem[\protect\citeauthoryear{Alam et~al.}{Alam et~al.}{2015}]{Alam:2015mbd}
Alam S.,  et~al., 2015, \mn@doi [Astrophys. J. Suppl.]
  {10.1088/0067-0049/219/1/12}, 219, 12

\bibitem[\protect\citeauthoryear{Alam et~al.}{Alam et~al.}{2017}]{Alam:2016hwk}
Alam S.,  et~al., 2017, \mn@doi [Mon. Not. Roy. Astron. Soc.]
  {10.1093/mnras/stx721}, 470, 2617

\bibitem[\protect\citeauthoryear{Alcock \& Paczynski}{Alcock \&
  Paczynski}{1979}]{Alcock:1979mp}
Alcock C.,  Paczynski B.,  1979, \mn@doi [Nature] {10.1038/281358a0}, 281, 358

\bibitem[\protect\citeauthoryear{Anderson et~al.}{Anderson
  et~al.}{2013}]{Anderson:2012sa}
Anderson L.,  et~al., 2013, \mn@doi [Mon. Not. Roy. Astron. Soc.]
  {10.1111/j.1365-2966.2012.22066.x}, 427, 3435

\bibitem[\protect\citeauthoryear{Anderson et~al.}{Anderson
  et~al.}{2014a}]{Anderson:2013oza}
Anderson L.,  et~al., 2014a, \mn@doi [Mon. Not. Roy. Astron. Soc.]
  {10.1093/mnras/stt2206}, 439, 83

\bibitem[\protect\citeauthoryear{Anderson et~al.}{Anderson
  et~al.}{2014b}]{Anderson:2013zyy}
Anderson L.,  et~al., 2014b, \mn@doi [Mon. Not. Roy. Astron. Soc.]
  {10.1093/mnras/stu523}, 441, 24

\bibitem[\protect\citeauthoryear{Beutler et~al.,}{Beutler
  et~al.}{2011}]{Beutler:2011hx}
Beutler F.,  et~al., 2011, \mn@doi [Mon. Not. Roy. Astron. Soc.]
  {10.1111/j.1365-2966.2011.19250.x}, 416, 3017

\bibitem[\protect\citeauthoryear{Blake et~al.}{Blake
  et~al.}{2011a}]{Blake:2011en}
Blake C.,  et~al., 2011a, \mn@doi [Mon. Not. Roy. Astron. Soc.]
  {10.1111/j.1365-2966.2011.19592.x}, 418, 1707

\bibitem[\protect\citeauthoryear{Blake et~al.}{Blake
  et~al.}{2011b}]{Blake:2011ep}
Blake C.,  et~al., 2011b, \mn@doi [Mon. Not. Roy. Astron. Soc.]
  {10.1111/j.1365-2966.2011.19606.x}, 418, 1725

\bibitem[\protect\citeauthoryear{Chuang \& Wang}{Chuang \&
  Wang}{2012}]{Chuang:2011fy}
Chuang C.-H.,  Wang Y.,  2012, \mn@doi [Mon. Not. Roy. Astron. Soc.]
  {10.1111/j.1365-2966.2012.21565.x}, 426, 226

\bibitem[\protect\citeauthoryear{Chuang \& Wang}{Chuang \&
  Wang}{2013}]{Chuang:2012ad}
Chuang C.-H.,  Wang Y.,  2013, \mn@doi [Mon. Not. Roy. Astron. Soc.]
  {10.1093/mnras/stt357}, 431, 2634

\bibitem[\protect\citeauthoryear{Cole}{Cole}{2011}]{Cole:2011zh}
Cole S.,  2011, \mn@doi [Mon. Not. Roy. Astron. Soc.]
  {10.1111/j.1365-2966.2011.19093.x}, 416, 739

\bibitem[\protect\citeauthoryear{Cole et~al.}{Cole et~al.}{2005}]{Cole:2005sx}
Cole S.,  et~al., 2005, \mn@doi [Mon. Not. Roy. Astron. Soc.]
  {10.1111/j.1365-2966.2005.09318.x}, 362, 505

\bibitem[\protect\citeauthoryear{Cuesta et~al.}{Cuesta
  et~al.}{2016}]{Cuesta:2015mqa}
Cuesta A.~J.,  et~al., 2016, \mn@doi [Mon. Not. Roy. Astron. Soc.]
  {10.1093/mnras/stw066}, 457, 1770

\bibitem[\protect\citeauthoryear{Eisenstein \& Hu}{Eisenstein \&
  Hu}{1998}]{Eisenstein:1997ik}
Eisenstein D.~J.,  Hu W.,  1998, \mn@doi [Astrophys. J.] {10.1086/305424}, 496,
  605

\bibitem[\protect\citeauthoryear{Eisenstein et~al.}{Eisenstein
  et~al.}{2005}]{Eisenstein:2005su}
Eisenstein D.~J.,  et~al., 2005, \mn@doi [Astrophys. J.] {10.1086/466512}, 633,
  560

\bibitem[\protect\citeauthoryear{Eisenstein, Seo  \& White}{Eisenstein
  et~al.}{2007a}]{Eisenstein:2006nj}
Eisenstein D.~J.,  Seo H.-j.,   White M.~J.,  2007a, \mn@doi [Astrophys. J.]
  {10.1086/518755}, 664, 660

\bibitem[\protect\citeauthoryear{Eisenstein, Seo, Sirko  \& Spergel}{Eisenstein
  et~al.}{2007b}]{Eisenstein:2006nk}
Eisenstein D.~J.,  Seo H.-j.,  Sirko E.,   Spergel D.,  2007b, \mn@doi
  [Astrophys. J.] {10.1086/518712}, 664, 675

\bibitem[\protect\citeauthoryear{Feldman, Kaiser  \& Peacock}{Feldman
  et~al.}{1994}]{Feldman:1993ky}
Feldman H.~A.,  Kaiser N.,   Peacock J.~A.,  1994, \mn@doi [Astrophys. J.]
  {10.1086/174036}, 426, 23

\bibitem[\protect\citeauthoryear{Fisher, Scharf  \& Lahav}{Fisher
  et~al.}{1994}]{Fisher:1993pz}
Fisher K.~B.,  Scharf C.~A.,   Lahav O.,  1994, \mn@doi [Mon. Not. Roy. Astron.
  Soc.] {10.1093/mnras/266.1.219}, 266, 219

\bibitem[\protect\citeauthoryear{Gaztanaga, Cabre  \& Hui}{Gaztanaga
  et~al.}{2009}]{Gaztanaga:2008xz}
Gaztanaga E.,  Cabre A.,   Hui L.,  2009, \mn@doi [Mon. Not. Roy. Astron. Soc.]
  {10.1111/j.1365-2966.2009.15405.x}, 399, 1663

\bibitem[\protect\citeauthoryear{Hartlap, Simon  \& Schneider}{Hartlap
  et~al.}{2006}]{Hartlap:2006kj}
Hartlap J.,  Simon P.,   Schneider P.,  2006, \mn@doi [Astron. Astrophys.]
  {10.1051/0004-6361:20066170}, 464, 399

\bibitem[\protect\citeauthoryear{Kaiser}{Kaiser}{1987}]{Kaiser:1987qv}
Kaiser N.,  1987, Mon. Not. Roy. Astron. Soc., 227, 1

\bibitem[\protect\citeauthoryear{Kazin, Sanchez  \& Blanton}{Kazin
  et~al.}{2012}]{Kazin:2011xt}
Kazin E.~A.,  Sanchez A.~G.,   Blanton M.~R.,  2012, \mn@doi [Mon. Not. Roy.
  Astron. Soc.] {10.1111/j.1365-2966.2011.19962.x}, 419, 3223

\bibitem[\protect\citeauthoryear{Kazin et~al.}{Kazin
  et~al.}{2013}]{Kazin:2013rxa}
Kazin E.~A.,  et~al., 2013, \mn@doi [Mon. Not. Roy. Astron. Soc.]
  {10.1093/mnras/stt1261}, 435, 64

\bibitem[\protect\citeauthoryear{Landy \& Szalay}{Landy \&
  Szalay}{1993}]{Landy:1993yu}
Landy S.~D.,  Szalay A.~S.,  1993, \mn@doi [Astrophys. J.] {10.1086/172900},
  412, 64

\bibitem[\protect\citeauthoryear{Lewis, Challinor  \& Lasenby}{Lewis
  et~al.}{2000}]{Lewis:1999bs}
Lewis A.,  Challinor A.,   Lasenby A.,  2000, \mn@doi [Astrophys. J.]
  {10.1086/309179}, 538, 473

\bibitem[\protect\citeauthoryear{Okumura, Matsubara, Eisenstein, Kayo, Hikage,
  Szalay  \& Schneider}{Okumura et~al.}{2008}]{Okumura:2007br}
Okumura T.,  Matsubara T.,  Eisenstein D.~J.,  Kayo I.,  Hikage C.,  Szalay
  A.~S.,   Schneider D.~P.,  2008, \mn@doi [Astrophys. J.] {10.1086/528951},
  676, 889

\bibitem[\protect\citeauthoryear{Padmanabhan \& White}{Padmanabhan \&
  White}{2008}]{Padmanabhan:2008ag}
Padmanabhan N.,  White M.~J.,  2008, \mn@doi [Phys. Rev.]
  {10.1103/PhysRevD.77.123540}, D77, 123540

\bibitem[\protect\citeauthoryear{Padmanabhan, Xu, Eisenstein, Scalzo, Cuesta,
  Mehta  \& Kazin}{Padmanabhan et~al.}{2012}]{Padmanabhan:2012hf}
Padmanabhan N.,  Xu X.,  Eisenstein D.~J.,  Scalzo R.,  Cuesta A.~J.,  Mehta
  K.~T.,   Kazin E.,  2012, \mn@doi [Mon. Not. Roy. Astron. Soc.]
  {10.1111/j.1365-2966.2012.21888.x}, 427, 2132

\bibitem[\protect\citeauthoryear{Park, Vogeley, Geller  \& Huchra}{Park
  et~al.}{1994}]{Park:1994fa}
Park C.,  Vogeley M.~S.,  Geller M.~J.,   Huchra J.~P.,  1994, \mn@doi
  [Astrophys. J.] {10.1086/174508}, 431, 569

\bibitem[\protect\citeauthoryear{Reid et~al.}{Reid et~al.}{2016}]{Reid:2015gra}
Reid B.,  et~al., 2016, \mn@doi [Mon. Not. Roy. Astron. Soc.]
  {10.1093/mnras/stv2382}, 455, 1553

\bibitem[\protect\citeauthoryear{Ross et~al.}{Ross et~al.}{2012}]{Ross:2012qm}
Ross A.~J.,  et~al., 2012, \mn@doi [Mon. Not. Roy. Astron. Soc.]
  {10.1111/j.1365-2966.2012.21235.x}, 424, 564

\bibitem[\protect\citeauthoryear{Taruya, Saito  \& Nishimichi}{Taruya
  et~al.}{2011}]{Taruya:2011tz}
Taruya A.,  Saito S.,   Nishimichi T.,  2011, \mn@doi [Phys. Rev.]
  {10.1103/PhysRevD.83.103527}, D83, 103527

\bibitem[\protect\citeauthoryear{Vargas-Maga\~{n}a, Ho, Fromenteau  \&
  Cuesta}{Vargas-Maga\~{n}a et~al.}{2015}]{Vargas-Magana:2015rqa}
Vargas-Maga\~{n}a M.,  Ho S.,  Fromenteau S.,   Cuesta A.~J.,  2015, preprint (arXiv:1509.06384)

\bibitem[\protect\citeauthoryear{Vargas-Maga\~{n}a et~al.}{Vargas-Maga\~{n}a
  et~al.}{2016}]{Vargas-Magana:2016imr}
Vargas-Maga\~{n}a M.,  et~al., 2016, preprint (arXiv:1610.03506) 

\bibitem[\protect\citeauthoryear{White, Tinker  \& McBride}{White
  et~al.}{2014}]{White:2013psd}
White M.,  Tinker J.~L.,   McBride C.~K.,  2014, \mn@doi [Mon. Not. Roy.
  Astron. Soc.] {10.1093/mnras/stt2071}, 437, 2594

\bibitem[\protect\citeauthoryear{Xu, Cuesta, Padmanabhan, Eisenstein  \&
  McBride}{Xu et~al.}{2013}]{Xu:2012fw}
Xu X.,  Cuesta A.~J.,  Padmanabhan N.,  Eisenstein D.~J.,   McBride C.~K.,
  2013, \mn@doi [Mon. Not. Roy. Astron. Soc.] {10.1093/mnras/stt379}, 431, 2834

\makeatother
\end{thebibliography}

\bsp	% typesetting comment
\label{lastpage}
\end{document}